\documentclass[times,authoryear]{elsarticle}

\usepackage{jasr}
\usepackage{framed,multirow}
\usepackage{tabulary}
\usepackage{amssymb}
\usepackage{latexsym}
\usepackage{adjustbox}

\usepackage[switch]{lineno}

\usepackage{url}
\usepackage{xcolor}
\definecolor{newcolor}{rgb}{.8,.349,.1}

\usepackage[citebordercolor=white]{hyperref}

\journal{Advances in Space Research}

\begin{document}

\verso{Ankush Bhaskar \textit{et.al.}}

\begin{frontmatter}
\title{\textbf{AuroraMag}: Twin Explorer of Asymmetry in Aurora and Solar Wind-Magnetosphere Coupling}


\author[1]{Ankush  \snm{Bhaskar}\corref{cor1}}
\cortext[cor1]{Space Physics laboratory, ISRO/VSSC,  India}
\ead{ankush_bhaskar@vssc.gov.in}
\author[1]{Jayadev \snm{ Pradeep}}
\author[2]{Shyama \snm{ Narendranath}}
\author[5]{ Dibyendu  \snm{ Nandy}}
\author[4]{ Bhargav \snm{ Vaidya}}
\author[7]{ Priyadarshan \snm{  Hari}}
\author[1]{ Smitha V. \snm{ Thampi}}
\author[1]{ Vipin K. \snm{ Yadav}}
\author[3]{ Geeta \snm{ Vichare}}
\author[6]{ Anil  \snm{ Raghav}}
\author[8]{ Dibyendu   \snm{ Chakrabarty}}
\author[1]{ R. Satheesh \snm{ Thampi }}
\author[1]{ Tarun Kumar \snm{Pant}}

\affiliation[1]{organization={Space Physics Laboratory, Vikram Sarabhai Space Centre, ISRO },
                city={Trivandrum},
                postcode={},
                country={India}}

\affiliation[2]{organization={ U. R. Rao Satellite Centre, ISRO },
                city={Bengaluru},
                postcode={},
                country={India}}
                
\affiliation[3]{organization={ Indian Institute of Geomagnetism },
                city={Mumbai},
                country={India}}

\affiliation[4]{organization={ Indian Institute Technology- Indore },
                city={Indore},
                country={India}}

\affiliation[5]{organization={Center of Excellence in Space Sciences India, Indian Institute of Science Education and Research Kolkata },
                city={Kolkata},
                country={India}}
\affiliation[6]{organization={Depratment of Physiscs, University of Mumbai },
                city={Mumbai},
                country={India}}
                
\affiliation[7]{organization={ Indian Institute of Space Science and Technology },
                city={Trivandrum},
                country={India}}   

\affiliation[8]{organization={ Physical Research Laboratory },
                city={Ahamdabad},
                postcode={},
                country={India}} 
\received{}
\finalform{}
\accepted{}
\availableonline{}

\begin{abstract}
In the present-day context, small satellites and their constellations consisting of varying sizes (nano, micro, pico satellites) are being favored for remote sensing and in situ probing of the heliosphere and terrestrial magnetosphere-ionosphere system. We introduce a mission concept aimed at concurrently observing Earth's northern and southern auroral ovals while conducting in situ measurements of particles, fields, and temperature. The mission concept consists of two small satellites, each having an identical auroral X-ray imager, an in situ particle detector, a magnetometer pair, and an electron temperature analyzer onboard in an elliptical polar orbit (400X1000 km ).  This mission would assist the space weather community in primarily answering important questions about the formation, morphology, and hemispherical asymmetries that we observe in the X-ray aurora, the fluxes of precipitating particles, Solar Energetic Particles, currents, and cusp dynamics. Once realized, this would be the first dedicated twin spacecraft mission of such kind to simultaneously study hemispheric asymmetries of solar-wind magnetosphere coupling. This study reveals the intricacies of the mission concept, encompassing orbital details, potential payloads, and its underlying scientific objectives. By leveraging the capabilities of small satellites, this mission concept is poised to make significant contributions to space weather monitoring and research.

\end{abstract}


\end{frontmatter}


\section{Introduction}
\label{sec1}

\begin{figure}
    \centering
    \includegraphics[width=0.6\linewidth]{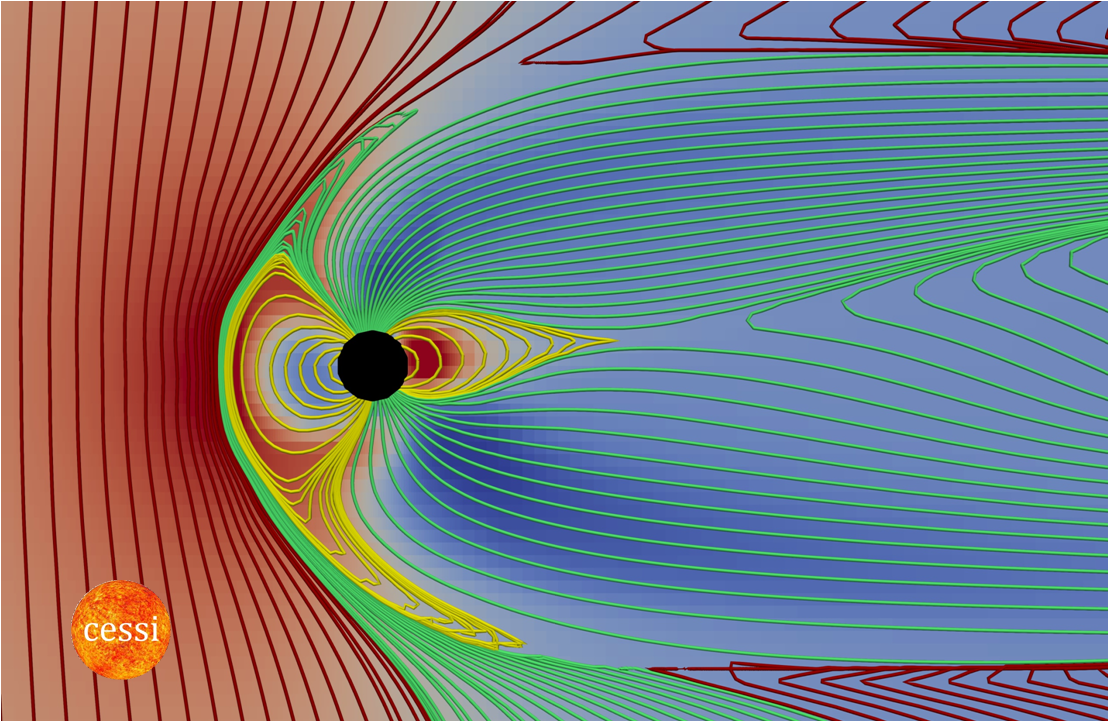}
    \caption{The global shape of the magnetosphere is very well depicted by the Magnetohydrodynamic simulation. Also, it clearly shows local time asymmetries in the shape and configuration of the magnetosphere. Magnetic fields are depicted by curved lines, with deep red, yellow, and green depicting solar wind, magnetosphere, and reconnected field lines, respectively.}
    \label{fig:magneto}
\end{figure}

Understanding and predicting space weather has become of paramount importance in the current space era as the number of satellites will continue to increase along with the usage of space \citep{virgili2016risk}. The continuous stream of particles originating from the Sun, known as Solar wind, compresses the Earth's magnetic cavity, forming a comet-like structure.  Figure ~\ref{fig:magneto} shows the Magnetohydrodynamic simulation of the solar wind-forced space environment of the Earth and reveals the typical global-scale configuration of the steady state magnetosphere \citep{das2019modeling}. The supersonic solar wind carrying solar magnetic fields impacts the Earth's magnetosphere from the day side (left of image), forming a bow shock structure around the magnetopause which plays a crucial role in shielding the inner atmosphere from the solar wind. On the night side, magnetic reconnection and dynamics induced by the solar wind lead to the formation of an extended magnetotail and a current sheet.

The solar wind magnetosphere coupling is the key to understanding and forecasting the influence of solar wind on the Earth's space environment.  The colorful displays in the northern and southern high latitude region, known as Aurorae are an excellent manifestation of this coupling. Humans have observed auroras since the Stone Age, but scientific study has only developed in recent centuries(for example \cite{eather1980majestic, hayakawa2015records,bhaskar2020analysis}).

 Aurorae have been observed on several planets besides Earth, including Jupiter and Saturn \citep{bhardwaj2000auroral,tao2011uv}. Ever since the dawn of the space era, aurora has been observed routinely by space-based imagers like imagers onboard NOAA and DMSP satellites, along with ground-based networks in high latitude region\citep{frank1988imaging,luhr1998westward,chisham2007decade,burch2000image}. After decades of research using space and ground-based platforms, there remain several unresolved questions that require investigation to better understand solar wind-magnetosphere-ionosphere coupling. One of the key features of aurora is its hemispherical asymmetry. Studying the asymmetry of auroras in the Northern and Southern Hemispheres is important in the field of space weather and planetary science. There are only a handful of reports \citep{laundal2009asymmetric,liou2019hemispheric} containing investigation on the auroral asymmetry. 

 There are some constraints to using ground-based simultaneous observations of both hemispheres have a narrow field of view and do not capture full auroral shape.  In the past, The Imager for Magnetopause-to-Aurora Global Explorer(IMAGE)-like missions had some sporadic coverage of simultaneous observations of the aurora \citep{laundal2009asymmetric,liou2019hemispheric}. Defense Meteorological Satellite Program (DMSP), has been observing aurora in UV along with in situ measurements of precipitatining electrons and ions and significantly enhanced our understanding of auroral dynamics \citep{gussenhoven1981dmsp,shiokawa1997multievent,zhou2003shock}. Not only auroral development and particle precipitation but hemispherical asymmetry has also not been studied in the context of global effects of penetration electric fields during storms and substorms. This is important as the interplay of field-aligned currents plays a crucial role in determining the global effects of penetration electric fields. It is known that the field-aligned currents close through the ionospheric currents in the polar region and through currents (e.g. tail current) in the magnetotail. However, there is no dedicated mission at present to address the hemispherical asymmetry and disproportionate global effects (e.g. \cite{hui2017contribution} of storms and substorms in a comprehensive manner.

There are several upcoming small satellite missions to probe the geospace environment \citep{miyashita2023brief}. Here we discuss a few of them to give some idea of the current scenario of geospace probing using smallsats.  Geostationary Transfer Orbit Satellite(GTOsat) is a 6U CubeSat mission from the USA \citep{blum2020gtosat}. The orbit of GTOsat is about 185 km $\times$ 5.5 $R_E$ in altitude with a low inclination of $< 25^{o}$.  GTOSat’s main targeted science is radiation belt dynamics. Another upcoming 12U Cubesat mission from Japan is going to capture global images of the magnetopause and cusps through soft X-ray emission, GEOspace X-Ray Imager (GEO-X) with an apogee of $\sim$ 40 $R_E$ \citep{ezoe2020geo}.
Magnetospheric Constellation (MagCon) from the USA, aims to study the energy storage and release processes in the magnetotail during substorms \citep{kepko2018magnetospheric}. In this constellation, there are 12 proposed spacecraft planned for distribution between 6-8 and 25 $R_E$ with a low inclination orbit. Each spacecraft will measure particles and DC magnetic fields. Self-Adaptive Magnetic Reconnection Explorer (AME) from China is intended to study magnetic reconnection at magnetopause and magnetotail by using one mother spacecraft and 12 CubeSats \citep{dai2020ame}. Ionospheric Neutron Spectrometer and Polar Explorer (INSPIRE) satellite is a microsat intended to probe ionospheric dynamics\citep{chandran2021inspiresat,evonosky2018inspiresat}. The number of proposed small satellite missions is steadily increasing, indicating a growing trend in such endeavors.


Following this brief overview, it is clear that there is presently no specific mission dedicated to exploring the hemispherical asymmetry of the aurora. Moreover, simultaneous imaging of aurora and particle measurements over the northern and southern hemispheres are currently not available from space-based platforms. Such measurements are very crucial to understanding the asymmetries in auroral formations and magnetospheric energy depositions in the southern and northern hemispheres.  The present mission concept, featuring twin spacecraft for the study of the aurora, builds upon the foundation laid in \cite{bhaskar2023aurora}. In the upcoming section, we delve into the pivotal scientific themes that guide the development of the proposed mission concept.

\section{ Driving Science of Hemispheric Asymmetries}
\subsection{Dynamics of X-ray Aurora and Asymmetry} 


\begin{figure}[ht!]
    \centering
    \includegraphics[width=0.9\linewidth]{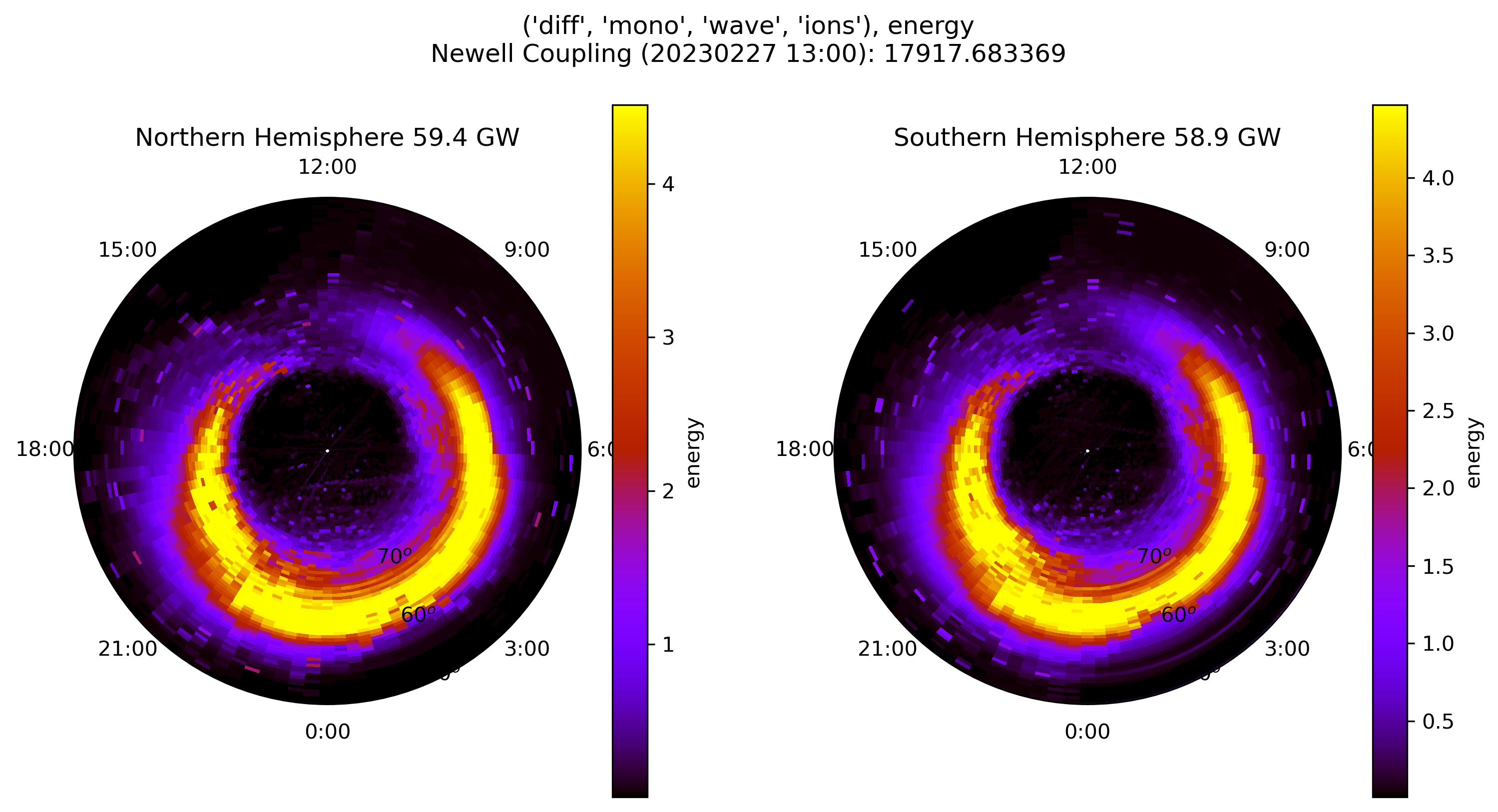}
    \caption{The figure shows modeled auroral energy deposits in the (left) northern and  (right) southern hemispheres using  OVATION Prime model \citep{newell2009diffuse}}
    \label{fig:aurora}
\end{figure}
Aurorae are ubiquitous phenomena with radiation emitted in a broad electromagnetic band. Magnetospheric electrons precipitate along the poleward field lines experiencing an acceleration in parallel electric fields and impacting the ionosphere leading to auroras. The visible aurora results from the excitation of atmospheric molecules such as N and O. X-rays are generated by the bremsstrahlung of the keV electrons as they precipitate towards the poles.  The precipitating accelerated electrons can be described by a Maxwellian distribution and by a power law at lower energies representing the pre-acceleration electrons. The X-ray spectrum generated by bremsstrahlung would depend on the initial energy distribution of the electrons and loss rates as they travel through the atmosphere. The X-ray image and spectra during aurora can be inverted to map the spatial distribution of the precipitating electrons and their spectra. Most of what is known about X-ray auroral emissions is from observations from the PIXIE instrument \citep{ostgaard2001auroral} on the POLAR satellite that was in operation from 1996 to 2003. Published results from these measurements are in the energy range of ~3 keV to 22 keV.  The only image of X-rays $< 3$ keV of the aurora is from a single observation of Earth by Chandra X-ray Observatory \citep{elsner2005simultaneous}. Here again, there are no spectroscopic measurements in this band. These X-ray measurements have shown variability at the time scale of minutes. The X-ray emission has a narrow structure within the arc with additional discrete emissions. The periodic space-based X-ray observation of the aurora will provide valuable data on the uncharted territory of X-ray emissions of the aurora.  

The interhemispheric asymmetry in aurora is very crucial to understanding the energy budget of the magnetosphere. Past observations from DMSP satellites in the UV band has helped to model the energy deposited in the magnetosphere. Also, the hemispheric asymmetry of aurora has been investigated using long-term ground-based observations from the Arctic and Antarctica \citep{hu2014hemispheric}.  Figure \ref{fig:aurora} shows the model output of OVATION Prime model \citep{newell2009diffuse} during the main phase of the geomagnetic storm on February 27, 2023. The auroral oval pattern is almost similar, but there is an asymmetry in deposited energy. Thus observing aurora in x-rays and simultaneously over both the auroral oval will provide first-of-its-kind measurements.


\subsection{Dynamics of Radiation Belts and Asymmetry in particle precipitation }
\begin{figure}
     \centering
     \includegraphics[width=0.9\linewidth]{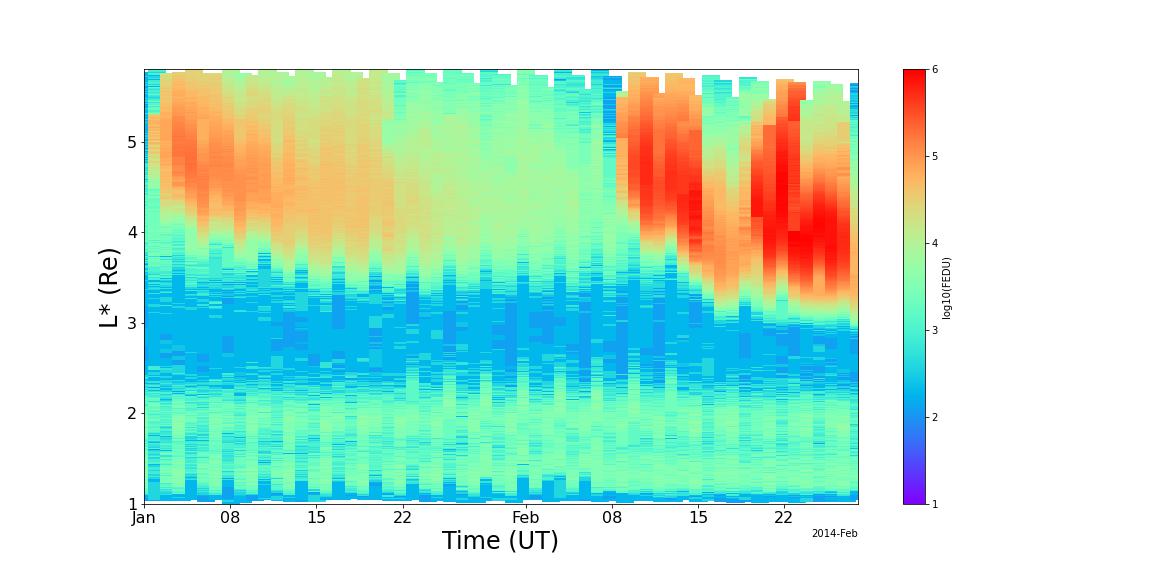}
     \caption{Radition belt variability as observed in MeV electrons measured by REPT onboard Van Allen Probe A.}
     \label{fig:rept}
 \end{figure}

Along with, remote sensing of the aurora, many space-based in situ observations of trapped and precipitating electrons have been continuously carried out, which has assisted in fathoming the energetics of the auroral particles. \citep{baker2012relativistic,reeves2003acceleration,roederer2012dynamics,schiller2016prompt}. The radiation belt harbors high-energy electrons and protons which are accelerated by various magnetospheric processes \citep{fok2008radiation,tsurutani1981wave,miyoshi2010time}. Specifically, the outer radiation belt is highly dynamic and varies from minute to solar cycle time scales. Figure \ref{fig:rept} shows the variability of relativistic 
 electrons of 1.8 MeV for different L-shells as measured by the Van Allen Probe A.  The variability of particle fluxes is the net effect of injection, energization, transport, and loss. Transient events like geomagnetic storms, interplanetary shocks, solar wind dynamic pressure changes, and bursts of wave activity cause sudden energization or losses of the particles within the magnetosphere \citep{shprits2008review,thorne2010radiation,baker2012relativistic,kanekal2016prompt,bhaskar2021radiation,chakraborty2022statistical}. The observatory class missions like NASA's Van Allen Probes \citep{spence2013science} and Exploration of energization and JAXA's Radiation in Geospace (ERG) \citep{miyoshi2018geospace} have immensely improved our understanding about the radiation belt source and sink processes.

The past or current missions were not specifically dedicated to measuring simultaneously particle fluxes precipitating in the southern and northern hemispheres. The asymmetry in particle deposition in the auroral regions is not well explored. Thus there is a need to have simultaneous in situ measurements of the precipitating particle fluxes along with imaging aurora which will give a better understanding of the sink and acceleration processes of the magnetospheric particles.

\subsection{Dynamics of Magnetospheric currents and Asymmetry } 
The dynamic and complex magnetospheric current systems include magnetopause current, magnetotail current, ring current in the outer radiation belt, substorm current wedge, field-aligned currents (FACs) connecting magnetosphere and ionosphere, DP2 current cells, etc. The eastward and westward electrojets flowing in the auroral zone are primarily the Hall currents and are confined to the auroral oval because of the enhanced ionization due to precipitating particles \citep{treumann1997advanced}. The eastward electrojet flows from afternoon to midnight through dusk and terminates at Harang discontinuity. The westward electrojet on the other hand flows through morning and midnight extending till the evening sector.
Due to the tilt of the Earth’s magnetic axis, differences in solar radiation introduce differences in the day-night conductivities in the polar region, which results in hemispheric differences in auroral intensities \citep{laundal2009asymmetric}. The increased radiation around the poles enhances the thermospheric response to external disturbances. In addition to seasons, solar wind impact angle, angle of inclination, IMF B$_y$ orientation, etc. give rise to the hemispheric asymmetries in the magnetospheric current systems. The potential patterns, high-latitude plasma convection, thermospheric density, and wind profiles are not symmetric in both hemispheres due to the IMF-By component \citep{cowley1981magnetospheric,forster2008high,ostgaard2011interhemispherical,reistad2016dynamic,tanaka2001interplanetary,yamazaki2015north,forster2015interhemispheric}. The effects of IMF B$_y$ on the penetration electric field (e.g. \cite{chakrabarty2017role}) and ring current (e.g. \citep{kumar2020effects}) have also been brought out. These are important indications that IMF By can play an important role in generating the global asymmetry of the effects of space weather events. 

The compression of the magnetospheric cavity due to impinging solar wind with an angle twist, distorts the magnetic field geometry, displacing the footpoints of the magnetic field lines in both hemispheres. One such clue can be obtained from the dependence of the effects of the penetration of electric fields on the solar wind azimuthal angle as shown by \cite{rout2017solar}. These could result in the latitudinal differences in the occurrence of the aurora, Region 1, and Region 2 currents, and their MLT patterns. 

\begin{figure} [ht!]
    \centering
    \includegraphics[width=0.75\linewidth]{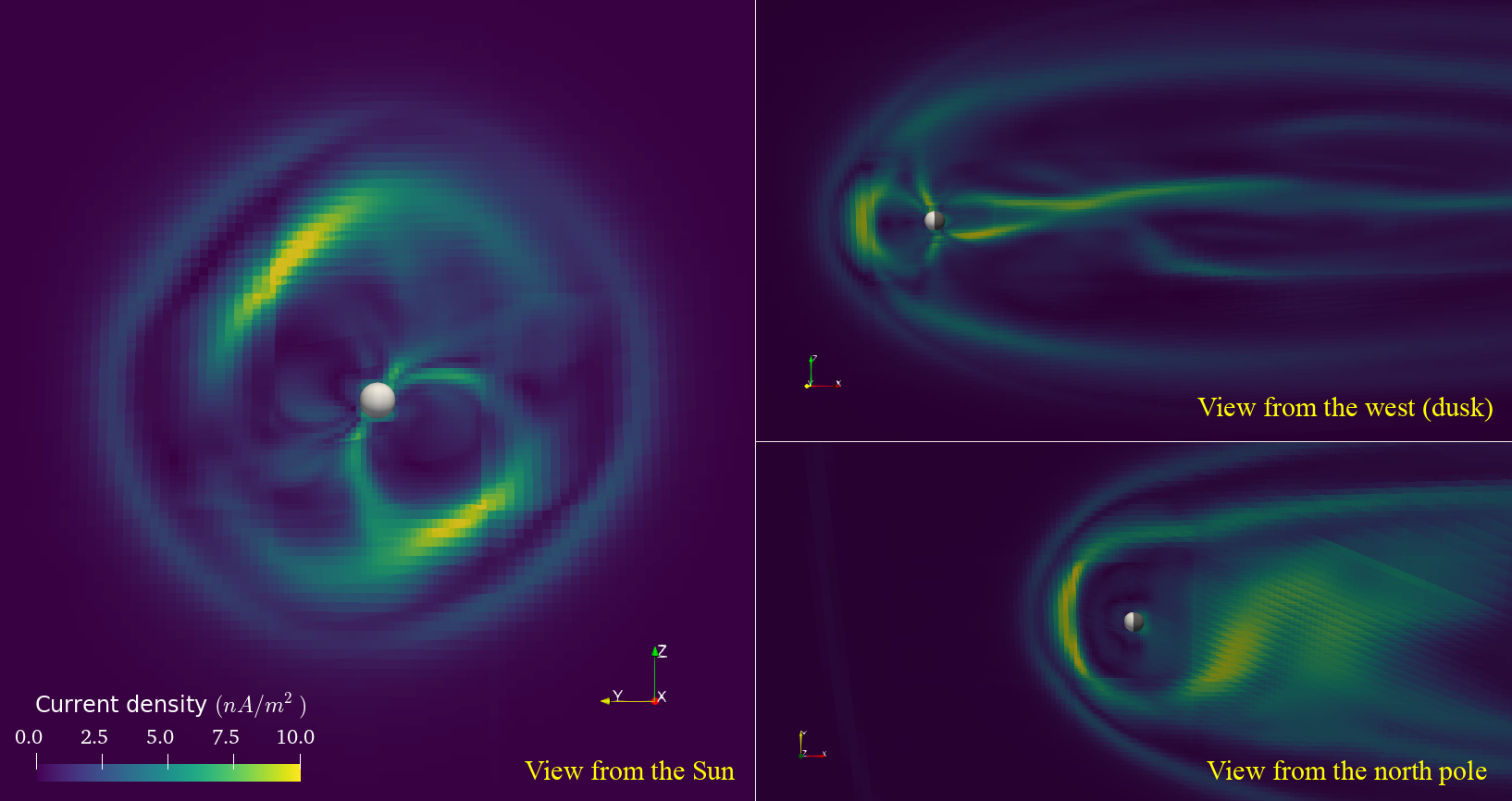}
    \caption{Simulated magnetosphere currents for the impact of ICME \citep{roy2023time}. The left panel shows a view from the Sun-side to a plane parallel to the Earth's dipole axis. The top-right panel shows a view from the west (dusk) of a plane normal to the equatorial plane. The bottom-right panel portrays a view from the north pole to the equatorial plane. The day and night hemispheres of the Earth are colored with white and black, respectively.}
    \label{fig:simucurrent}
\end{figure}

The impulses of solar wind pressure, ICMEs, substorms, and FACs introduce the asymmetries in the ring current. The partial ring current that couples the high and low latitude currents can also cause the hemispheric differences. Chapman-Ferraro current flows in the magnetopause region. The tail current is primarily described as an equatorial nightside westward current outside 6.6 RE that flows in the vicinity of stretched magnetic field lines, closes on the magnetopause, and is carried by particles with energy less than 20 keV. The neutral sheet current separates the north and south lobes in the magnetic tail. The magnetotail gets shifted due to solar wind and also can make the north and south lobes asymmetric, changing the dynamics in both hemispheres. The global (3D) view of the induced current density distribution in the magnetosphere following the impact of a coronal mass ejection as simulated by the CESSI-STORM module\citep{roy2023time}.  Figure ~\ref{fig:simucurrent} demonstrates the current enhancement in the various regions of the magnetosphere like the magnetopause region (dayside), polar cap regions, and the neutral sheet region in the magnetotail. This simulation represents an overall understanding of the anisotropic and asymmetric impact of the coronal mass ejection on the Earth's magnetospheric currents. 

The observations to date and simulation studies indicate there is a need to have simultaneous magnetic field measurements from both hemispheres, which will help to investigate these asymmetries in magnetospheric currents in a more comprehensive manner. 


\subsection{High Energetic Particle's Access to Earth and Asymmetry } 
\subsubsection{Solar Energetic Particles}

\begin{figure}[ht!]
    \centering
    \includegraphics[width=0.8\linewidth]{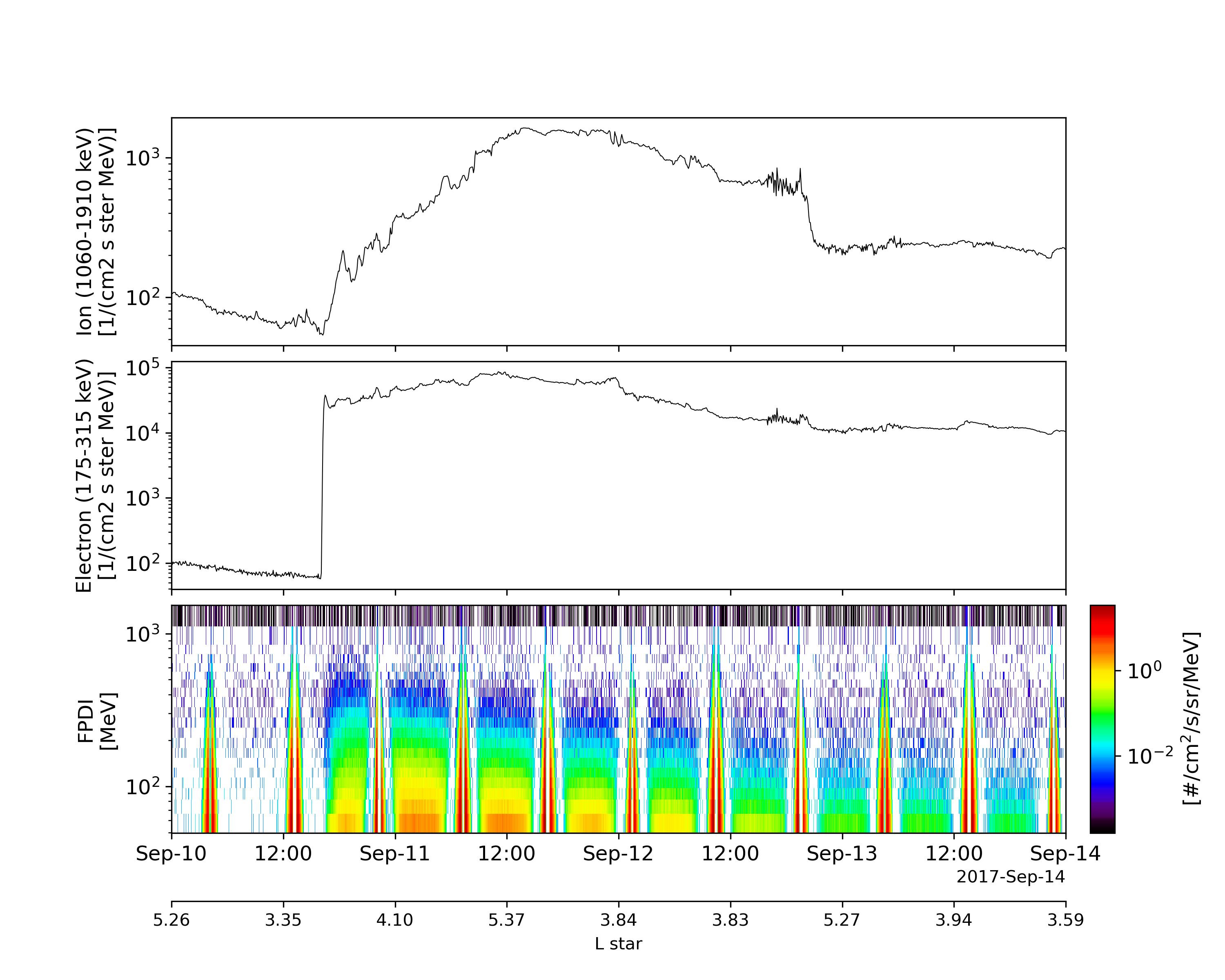}
    \caption{Solar Energetic Particle event of September 10, 2017, as observed by (top two panels) Advanced Composition Explorer (ACE) at L1 and (bottom panel) observed by Van Allen Probe -B inside the magnetosphere. }
    \label{fig:sep}
\end{figure}
The high-energy protons, electrons, and heavy ions of Solar Energetic Particles (SEPs) are accelerated to relativistic speeds by solar transient events including solar flares, ICMEs, and interplanetary shocks\citep{reames2013two,klein2017acceleration}. These particles are capable of penetrating deep inside the magnetosphere and affect satellites and even astronauts. The fraction of the penetrated population gets quasi-trapped and contributes to the inner irradiation belt \citep{hudson1995simulation,richard2009modeling,o2018solar, filwett2020solar, pandya2021quantitative}. The East-west anisotropy within the magnetosphere has been well studied using in situ measurements by various spacecraft such as Van Allen Probes \citep{o2018solar}. Figure ~\ref{fig:sep}  shows one of the strongest SEP events observed within the magnetopshere by The Relativistic Proton Spectrometer (RPS) instrument onboard Van Allen Probe-B. The sharp increase observed in electron and ion fluxes shown on top panels indicates the arrival of SEP at the L1 point also observed by the ACE spacecraft. After some time lag, we do see an increase in Proton differential Isotropic Flux (FPDI).  Due to asymmetry in the geomagnetic field, there is expected interhemispheric asymmetry in the penetration of these particles.  To investigate this asymmetry, trapping, and the propagation of SEPs within the magnetosphere simultaneous in situ measurements at both hemispheres will be of extraordinary importance.

\subsubsection{Origin of Ground Level Enhancements}
The sudden and short-lived sporadic increases in neutron detection have been occasionally observed by instruments placed to measure cosmic rays. These incidents were first documented by Forbush in 1942  and were subsequently termed "Ground Level Enhancements" (GLEs)\citep{forbush1950extraordinary}. Researchers have proposed that GLEs might be associated with solar energetic particles (SEPs). To date, there have been a total of 71 GLEs recorded since 1942, including 16 during Solar Cycle 23  \citep{cliver2006unusual, gopalswamy2012properties}. In general, it is assumed that extreme SEP events leading to GLEs are linked to solar flares and shocks caused by coronal mass ejections (CMEs). However, the origin of these GLEs remains a subject of debate \citep{aschwanden2012gev,firoz2014interpretation,klein2014relativistic,nitta2012special}. GLEs are a subset of SEP events that are observable on the Earth's surface, but many more energetic SEP events do not result in GLEs. Consequently, distinguishing between GLE events and SEP events can be somewhat ambiguous \citep{nitta2012special}.  

Researchers have thought that high-energy SEP particles interact with upper atmospheric nuclei and undergo spallation reactions, which produce numerous protons and neutron particles. While charged particles like protons interact electromagnetically with their surroundings along their path, neutrons, being charge-neutral, travel to the ground with minimal interaction. This leads to an observed increase in neutron flux at high-latitude neutron monitoring observatories. The observation of high-energy particles from ground and space-based platforms is critical in shedding light on the physical mechanisms underlying the formation of GLEs and the interconnections between SEP and GLE events.


\subsection{Dynamics of Cusp and Asymmetry } 
The Earth's cusp is a region where the solar wind particles interact directly with our planet. The Earth's cusp region has garnerated significant attention in space physics research due to its unique and dynamic nature. This transitional zone serves as a crucial entry point for solar wind particles and energy into the magnetosphere, playing a pivotal role in magnetospheric dynamics \citep{smith1996earth}. Recent studies utilizing satellite observations, ground-based measurements, and advanced modeling techniques have provided unprecedented insights into the complex plasma processes like magnetic reconnection, and particle acceleration occurring within the cusp \citep{fuselier2000cusp,chen1998cusp}. Furthermore, the cusp's variability and hemispheric asymmetric response to varying solar wind conditions have been explored to understand its impact on geomagnetic activity, auroral dynamics, and radiation belt dynamics \citep{kistler2010cusp,weygand2023interhemispheric,zhou2006cusp}. 

\subsubsection{Characterising FTE and Cusp interaction events}
\begin{figure}
    \includegraphics[width=1\columnwidth]{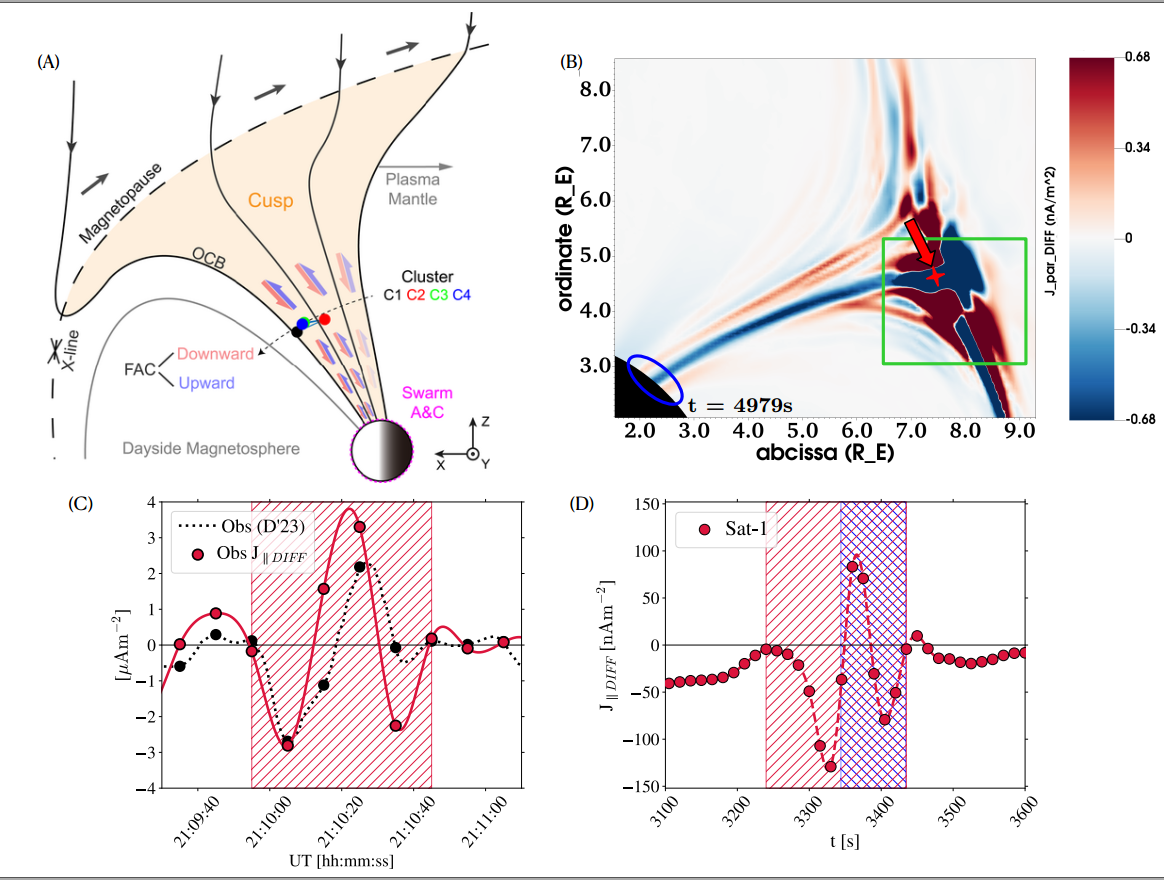}
\label{fig:FTECusp}
\caption{Panel (A) shows the cartoon of simultaneous observation of field-aligned currents (FACs) structure by Cluster and Swarm in the polar cusp region (adopted from \cite{Dong_2023}). The difference of FACs from MHD simulations using MagPIE \cite{Paul_2023} is shown in panel (B). The black dotted line in the panel 
(C) corresponds to an FAC observation by \cite{Dong_2023}. The black circles on the dotted curve represent the points where the data has been resampled. The solid red dots in panel (C) represent the temporal difference of FAC $J_{\parallel DIFF}$ obtained from the resampled observational data and the red solid line represents a spline fit to the scatter points. Panel (D) corresponds to the $J_{\parallel DIFF}$ obtained for simulated FTE Cusp interaction (see \cite{Paul_2023}}
\end{figure}

The Dungey convection cycle represents a fundamental phenomenon in space physics, wherein there is a transfer of energy and momentum from the solar wind to the magnetosphere, leading to circulation within the interconnected magnetosphere-ionosphere system. This cycle is observable through the movement of ionospheric plasma, which flows in an antisunward direction across the polar cap and then returns in a sunward direction just outside the polar cap within the auroral ovals, as described by \cite{Dungey_1961}. One of the primary processes that initiates this convection cycle is the development of an open magnetic field line due to magnetic reconnection via the solar wind southward $B_{z}$ component. The plasmoids generated during this process are typically referred to as Flux transfer events (FTEs) as they transfer the flux from the magnetosphere via the cusp interaction to the ionosphere. The FTEs and cusp interaction is primarily mediated by inter-change reconnection as shown in several MHD and hybrid-Vlasov simulations (see panel B figure~\ref{fig:FTECusp}) \citep[e.g.,][]{Lahti_2022, Paul_2023}. Such interactions in the dayside are crucial to understand as they channel both the Poynting flux and hydrodynamic energy flux from the Magnetosphere to the Ionosphere. Simultaneous measurements using the Cluster spacecraft and the SWARM satellites have reported such an event and the schematic diagram is shown in panel A of figure~\ref{fig:FTECusp} \citep{Dong_2023}. The inversion profile in the Filed-aligned Currnets (FACs )measured by SWARM (about 500 km orbit) is shown in panel C of the figure after a time delay of reconnection activity at the cusp detected by Cluster. Such a profile in FAC has also been reproduced by simulations as shown in panel D of the figure. 

The small satellites would be essential to provide additional evidences of such energy transfer events using simultaneous multi-point observations with existing magnetospheric missions like Magnetospheric Multiscale (MMS) and the upcoming Solar Wind Magnetosphere Ionosphere Link Explorer (SMILE). In addition, the effect of FTEs on inducing hemispheric asymmetry in the FAC would also be an important scientific aspect to target which will need twin smallsats simultaneously observing both hemispheres. Moreover, the in situ measurements of high-energy particles and ionospheric temperature are essential to link the high-energy protons injection due to acceleration via the interchange reconnections of FTEs with the cusp. 

\subsubsection{Cusp heating}
It is known that the electron temperature (T$_e$) in the F region of the ionosphere is determined by the competition between various heating, cooling, and energy flow processes \citep{schunk1978electron}. The heat gained by the ambient electron gas from the photoelectrons and from collisions with the neutrals increase the  T$_e$ above the ion and neutral temperatures. The hot ambient electrons then lose energy in Coulomb collisions with the ambient ions and in collisions with the neutral thermosphere. Therefore, the  T$_e$  depends on the relative importance of the various heating, cooling, and energy flow processes, and this in turn is dependent upon altitude, latitude, local time, season, solar activity, and space weather conditions. 
The electron energy balance is especially complicated at high latitudes owing primarily to the presence of additional heat sources \citep{schunk1978electron}, and actually, the T$_e$ morphology is indicative of cusp-like precipitation \citep{pallamraju2004first}. In the dayside cusp and the nocturnal auroral oval, precipitating energetic electrons are an important heat source for the ambient electron gas. Throughout the polar cap, convection electric fields having their source in the magnetosphere are responsible for a significant amount of Joule dissipation. In addition, field-aligned electron currents which are present at high latitudes cause further complications in electron energy balance considerations because of the need to consider thermoelectric energy transport.  The hemispherical asymmetry in cusp temperature due to the particle precipitation has been studied using DMSP observations \citep{newell1988hemispherical}. 
 
 This hemispheric asymmetry has profound implications for magnetospheric dynamics, plasma transport processes, and the overall response of the magnetosphere to solar wind variations. Understanding the cusp hemispheric asymmetry is crucial for deciphering the intricacies of magnetospheric interactions and their influence on space weather phenomena in different hemispheres. Therefore, continuous in situ measurements of temperature could assist in investigating various heating processes and quantifying them.


\subsection{Space Weather Impact Monitoring}

As mentioned earlier, the high-latitude ionosphere gets incident by precipitating particles and joule heating via magnetospheric-ionospheric currents. During intense or severe geomagnetic storms the energy entering into the magnetosphere-ionosphere system increases by orders of magnitude. The particle precipitation and joule heating in high latitudes cause changes in lower ionosphere conductivities, increasing radio wave absorption, and changes in atmospheric composition. The transient space weather events do affect existing space-based objects like communication and navigation satellites, space debris, space stations, and even airplanes. Humans in high altitudes near-polar regions and astronauts in space can easily get affected by sudden increases in the radiation dose. The Indian Space Research Organization (ISRO) has recently launched its first solar mission to the L1 point. The mission reached L1 on January 6, 2024, and has begun scientific observations of solar wind and the Sun, alongside NASA's ACE and WIND missions. In light of this, currently, there is continuous monitoring of solar wind and the Sun. There is a need to have a magnetospheric mission that will observe both hemispheres of the Earth simultaneously and quantify solar wind energy entering the magnetosphere-ionosphere system. The twin smallsat mission to study solar wind-magnetosphere coupling is timely as space weather forecasting has become a crucial aspect of the current space era. Moreover, the missions to monitor the upstream solar wind, including ACE, WIND,  and Aditya-L1 will provide the required critical inputs on solar and solar wind conditions to evaluate the geoeffectiveness of solar transients. 






Given science motivation and space weather monitoring aspects, there is an urgent need to have small satellite missions to study the magnetospheric variability and space weather impact. Smallsat missions have a quick turnaround time which will help to realize missions promptly.  Currently, there is no dedicated ongoing space mission to explore asymmetry using simultaneous observations from both hemispheres, which demands a dedicated twin smallsat mission to observe and quantify hemispherical asymmetry in Auroral and solar wind-magnetosphere coupling. Here we propose a twin small satellite magnetospheric mission, "AuroraMag" to address the posed science problems and improve space weather forecasting.  The observations from this will assist the global community in resolving multiple science questions in the auroral dynamics, radiation belt physics, SEPs-GLEs, Cusp-solar wind interaction, and solar wind-magnetosphere coupling processes. The following sections present details on the mission concept, payloads, and mission requirements.

\section{Mission Concept and Primary Science objectives } 
\begin{figure}[ht!]
  \centering
  \includegraphics[scale=0.25]{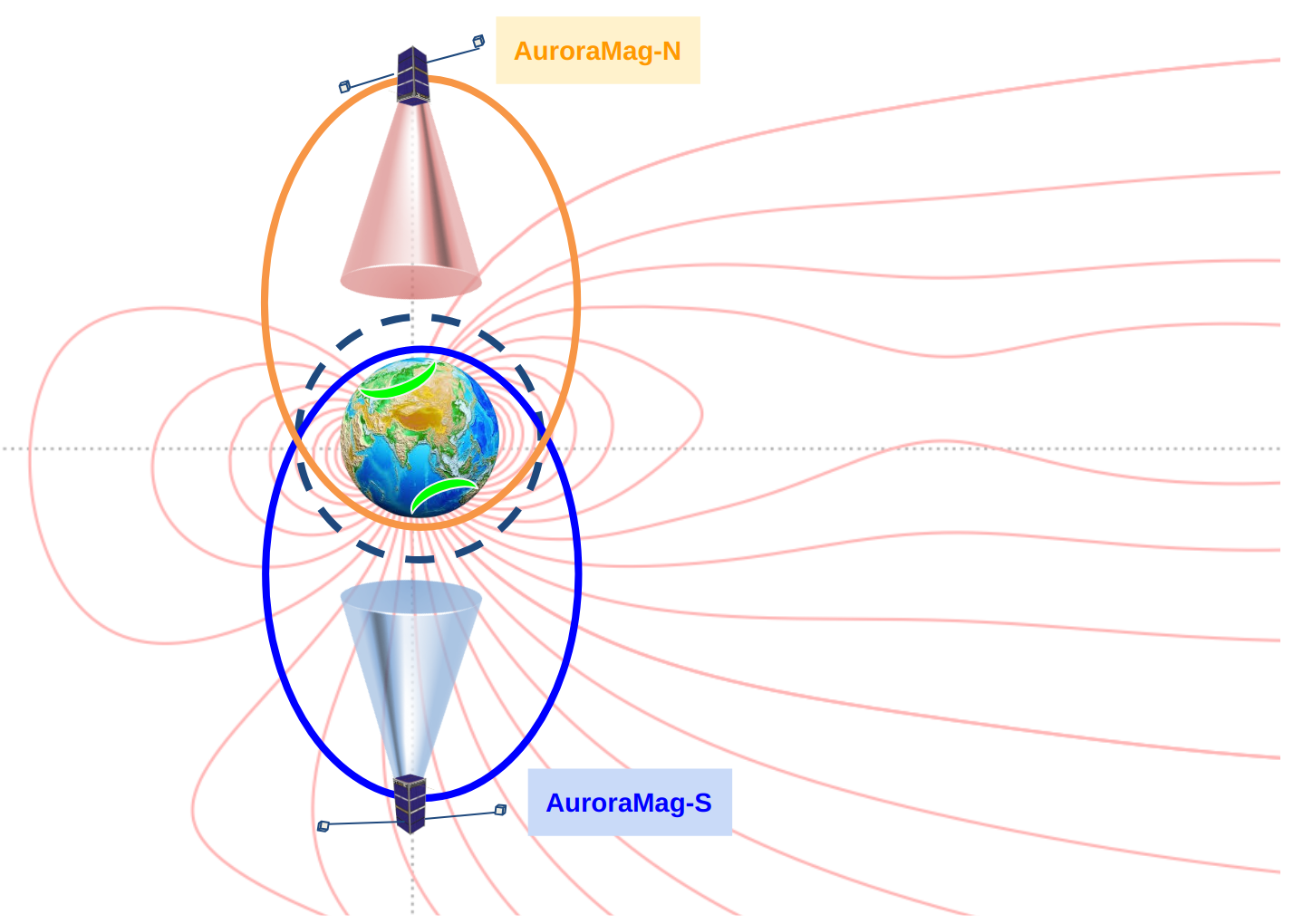}
  \caption{ Schematic of the \textit{AruroraMag} Twin smallsat mission concept (not to scale). AruroraMag-N will see a full northern auroral oval whereas AruroraMag-S will see a full southern auroral oval. A Black dashed circular orbit is shown for an alternative sun-synchronous orbit around 1000 km. }
  \label{fig:sketch}
\end{figure}

We propose two identical smallsats separated by optimal angular distance and having  $\sim 1000 $km Sun-Synchronous circular polar orbit. Figure ~\ref{fig:sketch} shows the cartoon of the twin smallsats observing both hemispheres simultaneously. The orbital parameters are optimized to get maximum auroral oval in the field-of-view (FOV) and longer simultaneous observation of it from both hemispheres. The orbital characteristic is discussed in more detail in the next section.

The following are the primary science objectives of the proposed mission concept. 

\begin{itemize}
    \item Investigate the hemispheric asymmetry of solar wind magnetosphere coupling in the southern and northern hemispheres during quiet and disturbed magnetospheric conditions  by simultaneous 
    \begin{itemize}
        \item imaging of aurora in X-ray.
        \item in situ measurements of  particle precipitation
        \item measurements of  magnetospheric/ionospheric currents
        \item in situ measurement of ionospheric electron density and temperature
    \end{itemize}

\item Estimate energy budget of the magnetosphere: Joule heating and Particle precipitation and their hemispherical asymmetric response.

\item Probe the high-latitude heating of the ionosphere and Cusp dynamics and, FTEs.

\item Determine the role of Magnetospheric waves in particle precipitation by combining observations from the spacecraft and ground-based observations

\end{itemize}

The science objectives, required instruments, and desired measurement specifications are summarized in Table 1. To address the identified science problems, we need an X-ray imager, particle detectors, electron temperature analyzers, and magnetometers onboard two spacecraft.

\begin{table}[h!]
\begin{tabulary}{\linewidth}{LLLL}
    \hline
      \textbf{Sr.} & \textbf{Science objective/Questions} & \textbf{Required Instrument} & \textbf{Measurement specification}  \\
          \hline
          \hline
         1& Interhemispheric Asymmetry in Auroral X-ray emissions and energy deposition  & X-ray Imager  &  Energy: 0.3 to 3 keV , FOV:$45^{\circ}$  \\
         2& Radiation belt dynamics and particle precipitation & X-ray Imager, MERiT, and Magnetometer &Precipitating electron fluxes, Electrons (5 keV to 10 MeV)   \\
          3& Cusp plasma and interaction with solar wind  & ETA & Electron Temperature ($^\circ K$)  \\
           4& Solar Energetic Particles entry into to Magnetosphere & MERiT & Protons (200 keV to 100 MeV)  \\
           5& Magnetospheric currents and interhemispheric asymmetry & Magnetometer & Vector magnetic field variations (accuracy $<1$ nT)    \\
          \hline
          \hline
\end{tabulary} 
\caption{Science Requirements Matrix}
\label{tab:ScReqMatrix}
\end{table}

\section{Orbital Geometry}

One of the critical aspects of designing a twin smallsat mission concept for simultaneous observations of Earth's northern and southern auroral ovals is the optimization of an orbital geometry that is conducive to meeting the specific science objectives of the mission. An ideal orbital configuration for the twin satellites should enable both the onboard X-ray imagers and \textit{in situ} instruments to have concurrent coverage over the auroral latitudes (typically poleward of 45° in terms of geographic latitudes) in both hemispheres, ensuring maximum favorable observational time in each orbit. 

The orbital geometry for the AuroraMag mission concept was optimized with the help of 3D orbital simulations implemented using the \textsc{poliastro} astrodynamics package \citep{rodriguez2016poliastro} of \textsc{python} for spacecraft orbital propagation, along with the \textsc{astropy} \citep{robitaille2013astropy} and \textsc{matplotlib} \citep{pajankar2022visualizing} libraries for vector algebra computations and 3D visualizations. A similar implementation of \textsc{python}-based 3D simulations of spacecraft-borne observational geometries has been extensively detailed by \cite{pradeep2023solar}, which was appropriately modified and adapted for the present work. Note that for simplicity, the simulations carried out in this work assume the Earth to be a perfect sphere having a radius equal to Earth's mean volumetric radius of 6371 km, and the due to this assumption is expected to be negligible for the purpose of this analysis. The rendered 3D simulations helped arrive at two optimized orbital configurations for AuroraMag: a primary elliptical configuration (400 km $\times$ 10000 km orbits), and an alternative circular configuration (1000 km orbit). 

\subsection{Elliptical Configuration: 400 km $\times$ 10000 km Orbits}

Primarily, a twin elliptical orbital configuration is envisaged for AuroraMag, in which the two smallsats (hereafter referred to as AuroraMag-N and AuroraMag-S) are placed into two separate polar elliptical orbits. The two orbits are identical in size and shape, but antipodal in terms of the location of their respective periapses. The rationale behind going for an elliptical orbit is to obtain coverage of a significant range of altitudes for \textit{in situ} profiling (using ETA, MERiT, and Magnetometer), while positioning the orbital apogee over the targeted auroral oval to maximize observation/imaging of the auroral latitudes. 

The orbits are optimized to a perigee altitude of 400 km and apogee altitude of 10000 km, which corresponds to a semimajor axis of 11571 km and eccentricity of 0.41483018. This comfortably enables comprehensive \textit{in situ} profiling of magnetospheric parameters over the 400$-$10000 km altitude range over each orbit. It was decided not to lower the perigee below 400 km as the effect of atmospheric drag at lower altitudes can lower the apoapsis over time, necessitating intermittent orbit corrections and diminishing the mission lifetime. The 10000 km apogee enables comfortable viewing of the full auroral ovals by the X-ray Imagers with a field-of-view of $\sim$45.8° ($\pm$22.9°), which corresponds to a latitudinal imaging extent of over 130° from the apogee. An even larger apoapsis is not preferred, considering possibly coarse spatial resolution for the imagers and increased launch costs.   

The two polar elliptical orbits are oriented antipodally so that the apoapses are over the respective target aurorae. Thus AuroraMag-N, which is designated to observe the northern auroral oval, is placed in an orbit with perigee over the South Pole (i.e., an argument of perigee of 270°). On the other hand, the orbit of AuroraMag-S has its perigee over the North Pole (i.e., argument of perigee of 90°) so that it covers the southern auroral oval on approaching apogee. The smallsats are intended to travel in opposite directions in their orbits, but synchronised to reach their respective apoapses and periapses at the same time, facilitating similar solar illumination conditions for the two spacecraft throughout. It is to be noted that since this proposed configuration of AuroraMag involves two orbits with diametrically opposite periapses and opposite-moving satellites, it may not be possible to achieve the mission through a single launch; two separate launches may be required to inject the twin smallsats into the indented orbits and achieve the stated objectives.

An inclination of 90° is adopted for both orbits to achieve coverage of the northern and southern auroral latitudes ($>$45° in either hemisphere). Given the direct correlation between the auroral ovals and the Earth-sun vector, both the satellites would be injected in an orbital plane perpendicular to the day-night terminator (i.e., noon-to-midnight orbits) for best coverage of auroral latitudes \& longitudes. It would have been ideal for the two smallsats to be in Sun-Synchronous Orbits (SSOs) in which case they would have maintained their orbital planes, but the significant ellipticity and large apoapsis of 10000 km implicate that it may not be feasible to achieve sun-synchronicity \citep[]{boain2004ab, liu2019guidance}. Thus, the two orbits in this configuration may undergo gravitational perturbations and gradually drift over time, and appropriate orbital correction maneuvers may be required to compensate for this. Nevertheless, these drifts may not significantly affect our observations given the large spatial coverage of the imager FOVs from the apoapses. Since we are not opting for SSOs, there are no additional constraints on the inclination and perfectly polar orbits (90° inclination) are adopted.

\begin{figure}
  \centering
  \includegraphics[scale=0.50]{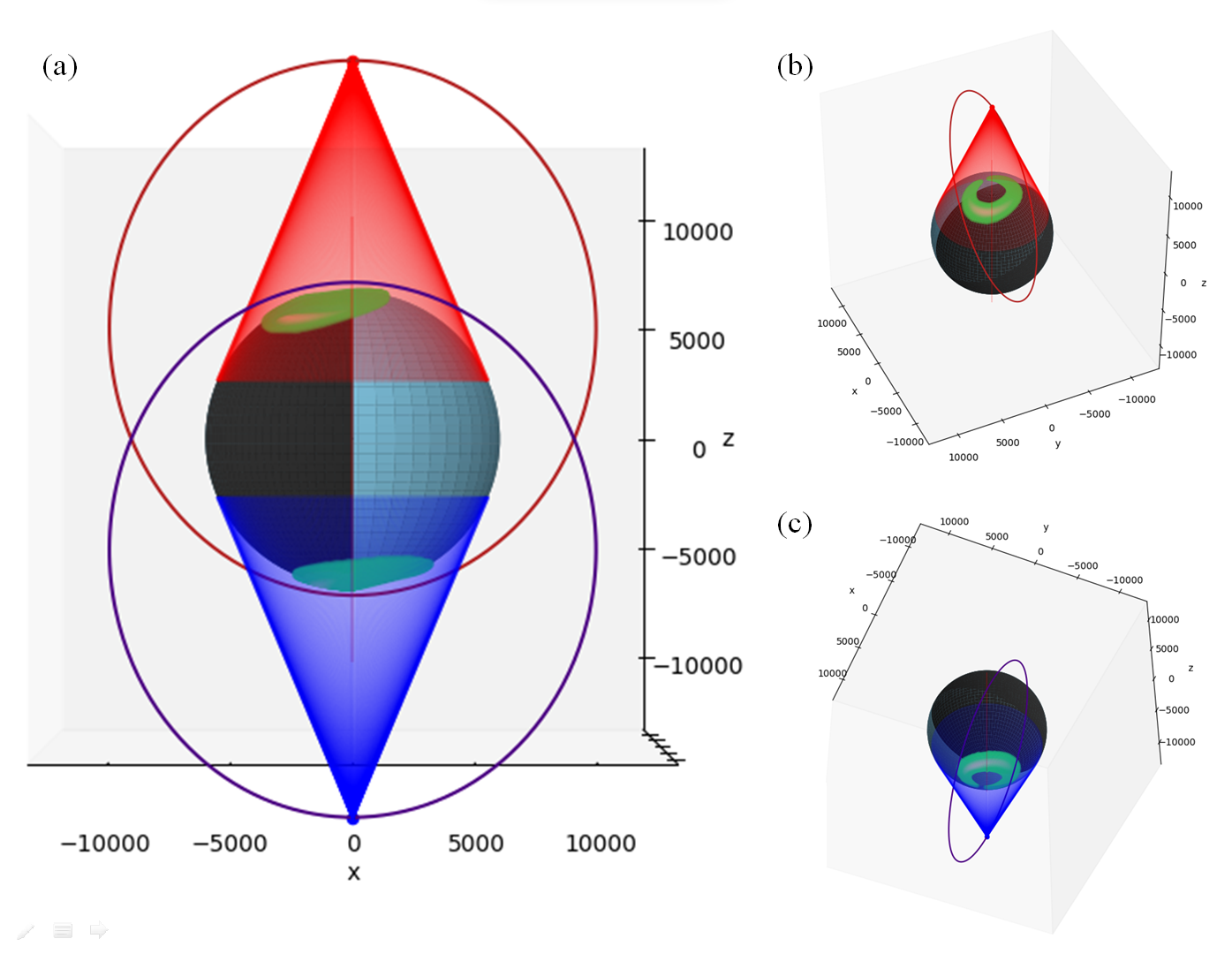}
  \caption{(a) 3D visualization of the optimized twin elliptical orbit configuration for AuroraMag mission concept, simulated around Earth ($blue$ sphere) as per the optimized parameters listed in Table \ref{tab:orb_opt}. The dark-shaded spherical region represents the night-side hemisphere of Earth under equinox conditions. The $red$ orbit corresponds to AuroraMag-N satellite and the $blue$ orbit corresponds to AuroraMag-S. The 3D $red$ and $blue$ 3D cones represent the 45.8° FOV of X-ray imagers onboard AuroraMag-N and AuroraMag-S, respectively. The auroral ovals are visualised in $green$ colour using data from  NOAA's OVATION Model (for the date 30 March 2023); (b) Simulated representation of northern auroral oval observation by AuroraMag-N; (c) Simulated representation of southern auroral oval observation by AuroraMag-S. The axes in each 3D plot correspond to Cartesian coordinates ($x$, $y$, and $z$), representing distance in kilometers (km).}
  \label{fig:orbitsim1}
\end{figure}

Incorporating the above considerations, 3D simulations were rendered to optimize the elliptical orbital configuration for AuroraMag. Table \ref{tab:orb_opt} summarises the optimized orbital parameters for the same. Figure \ref{fig:orbitsim1} presents 3D visualizations of the simulated twin elliptical orbital configuration of AurorMag around Earth, with the red and blue cones representing the FOVs of AuroraMag-N and  AuroraMag-S respectively, along with representative auroral data from the NOAA Ovation Model \citep{newell2014ovation} for a visual reference of our observational target.

\begin{table}[]
\centering
\begin{tabular}{c|c|c}
\textbf{Sl. No} & \textbf{Parameter}              & \textbf{Optimized Value}   \\[0.1cm] \hline 
1.              & Semimajor Axis                  & 11571 km \\[0.1cm]
2.              & Inclination                     & 90°               \\[0.1cm]
3.              & Eccentricity                    & 0.41483018                       \\[0.1cm]
4               & Orbital Period                     & $\sim$206.45 minutes             \\[0.1cm]
5.              & Argument of Periapsis & 270° (AuroraMag-N) \& 90°(AuroraMag-S)                       \\[0.1cm]
6.              & FOV of X-ray Imagers                     & 45.8° ($\pm$22.9°)                       \\[0.1cm]
\end{tabular}
\caption{Optimized parameters for the twin elliptical orbital configuration of AuroraMag smallsat mission concept.}
\label{tab:orb_opt}
\end{table}

\begin{figure}
  \centering
  \includegraphics[scale=0.35]{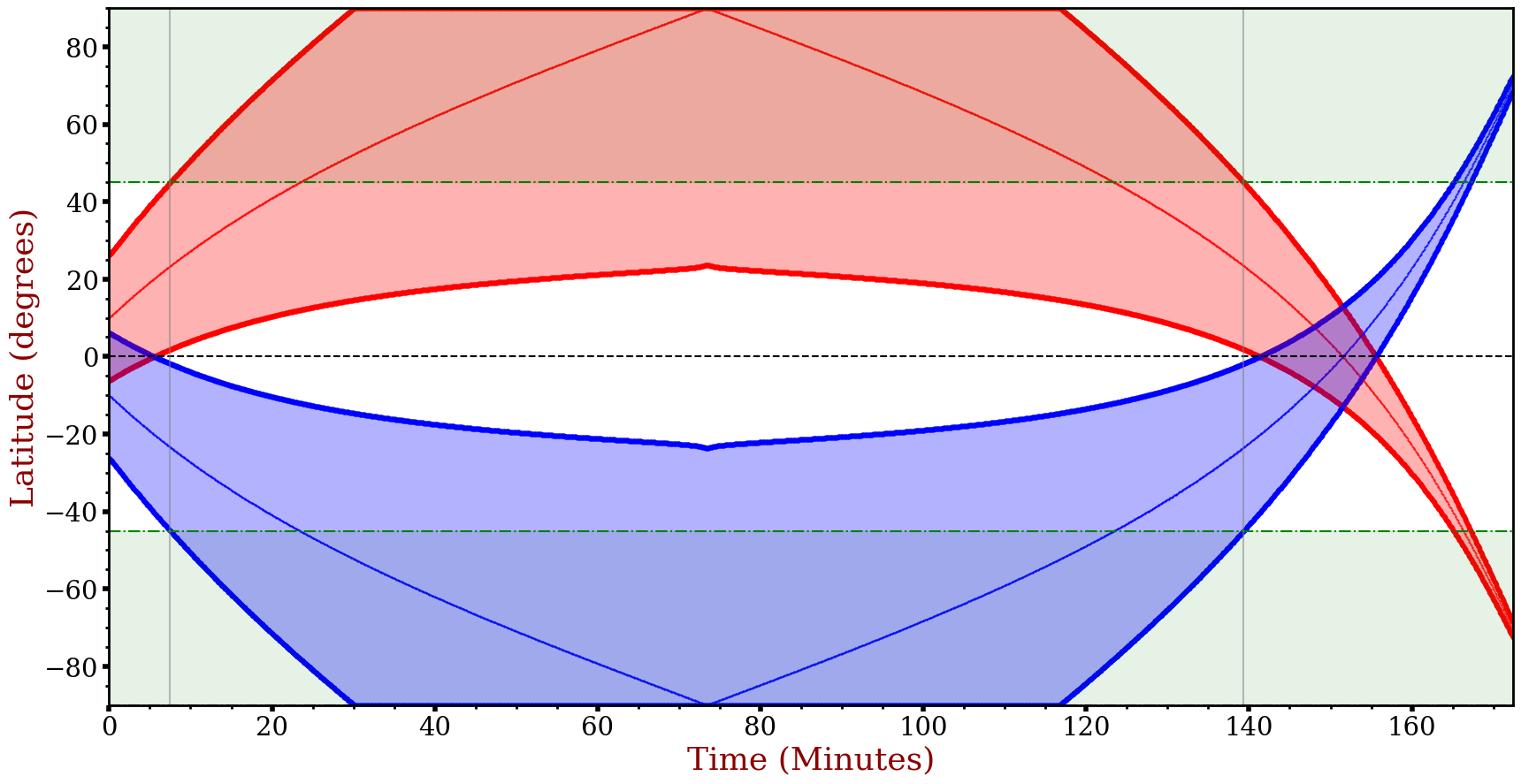}
  \caption{Results of 3D orbital simulations: time evolution of the FOV latitudinal coverage of AuroraMag X-ray imagers, simulated for the twin elliptical orbital configuration as per optimized parameters listed in \ref{tab:orb_opt}. The $y$-axis corresponds to Earth's geographical latitudes, while the $x$-axis denotes observation times, starting from an arbitrary point. The $red$ shaded region indicates the latitudinal coverage of the AuroraMag-N imager and the $blue$ shaded region indicates that of the AuroraMag-S imager. The thin $red$ curve corresponds to the latitudinal ground trace of AuroraMag-N, and the thin $blue$ curve corresponds to that of AuroraMag-S. The horizontal $green$-shaded regions indicate typical auroral latitudes in either hemisphere ($>$45$^{\circ}$ geographical latitude), and the vertical  $gray$ lines indicate the time bounds within which simultaneous observation of northern and southern auroral latitudes is achievable. }
  \label{fig:orbitsim_latcov}
\end{figure}

The \textsc{python}-based 3D simulations were invoked to determine the achievable duration of simultaneous observations of the auroral latitudes in each orbital period. For this, the two smallsats were simultaneously propagated over time along their respective simulated orbits using the in-built functions of \textsc{poliastro}. At each time step of the simulation, the fields of view of imagers onboard the two satellites (adopted as 45.8°) were visualized as 3-dimensional cones, and the intersection of these cones with the spherical surface of the planet (assumed to be a perfect sphere) was used to determine the spatial coverage of the observations over the Earth. It is to be noted that since we are employing X-ray Imagers rather than UV or visible wavelength detectors, our observations are not likely to be affected by scattered solar radiation on the sunlit hemisphere, enabling unhindered characterisation of the day-side aurora in addition to the night-side \citep{imhof1995polar}. The results of this simulation analysis are presented in Figure \ref{fig:orbitsim_latcov}. It is seen that for this configuration, we can achieve as long as around 132 minutes ($\sim$2.2 hours) of simultaneous observations of the auroral latitudes ($>$45° geographical latitude) over the two hemispheres, during every single orbital period (which is about 206.45 minutes or 3.44 hours for either orbit).


\subsection{Circular Configuration: 1000 km Orbit}

In addition to the primary elliptical orbital configuration, we also present the simulation results for an alternative circular orbit configuration for AuroraMag: a 1000 km noon-to-midnight sun-synchronous orbit for the twin smallsats. In this mission design, both AuroraMag-N and AuroraMag-S are placed into the same 1000 km near-polar orbit (semimajor axis or orbital radius of 7371 km), propagating in the same direction, and separated by an angle of 180° along the orbital plane. Given the circular orbit (eccentricity of 0) with 1000 km altitude, sun-synchronocity can be achieved in this case by adopting an orbital inclination of about 99.45$^{\circ}$ \citep{boain2004ab}, making the orbit near-polar and slightly retrograde to the Earth's rotation. 

This circular configuration provides a few advantages: (i) being a single-orbit twin satellite configuration, the two smallsats can be injected into orbit in a single launch, reducing the overall mission cost; (ii) the sun-synchronous orbit maintains orbital plane as required with minimal orbit corrections; (iii) both satellites can be used to study both Northern and Southern auroral ovals, giving significant favourable observational durations (around 84 minutes in every $\sim$104-minute orbit). However, this configuration also suffers from a couple of major drawbacks: (i) the circular orbit only offers a constant altitude (i.e., 1000 km) for \textit{in situ} measurements, making it impossible to obtain vertical profiles of magnetospheric parameters, unlike the previous elliptical configuration which facilitated profiling of the magnetosphere from 400 km to 10000 km; (ii) the lower orbital altitude also limits the spatial coverage of the imagers; for 1000 km altitude, a FOV 119.6° is feasible, which gives a latitudinal coverage just over 50°, which may not necessarily cover the entire auroral oval, giving only a partial picture. Owing to these drawbacks, we strongly recommend the twin elliptical orbital configuration for the AuroraMag mission concept, although the circular configuration can be considered a cost-effective alternative albeit with significantly diminished science returns.

The 3D orbital simulations enabled an interesting exploration of possible observational geometries for the AuroraMag twin smallsat mission concept. It should be noted that this exercise assumed a perfect sphericity for the Earth, ignoring the flattening of about 0.003353, but this is not expected to make any significant change in the results obtained. Atmospheric refraction effects and the actual altitude of occurrence of auroral emissions (around 300 to 400 km) have not been incorporated in the observational geometry computations carried out, but these are not likely to have a significant impact on the present analysis. The auroral data visualized in Figures \ref{fig:orbitsim1} should only be treated as a representative picture, as the data has been taken from the NOAA Ovation Model forecast for a single date (30 March 2023). 

\section{Payloads}
\subsection{Aurora X-ray Imager}

A wide coverage of the auroral region with high cadence, imaging, and spectroscopy would be the most ideal scenario. The X-ray imager aims to map the electrons precipitating into the ionosphere by measuring their bremsstrahlung emission in X-rays. Though the emission extends to hundreds of keV under high geomagnetic activity, we focus on the soft X-ray region where the studies are limited to isolated short observations from balloons and Astronomy satellites. This is a niche wavelength region that is also a link to the Vis-UV auroras and their high-energy counterparts. Additionally, a component of the solar wind charge exchange emission (SWCX) that plays an important role in the loss of atoms from the exosphere is primarily in the soft X-ray band. With the SMILE mission operational in $\sim 2025$, global views of the SWCX emission would be possible and a closer view by the X-ray imager would provide a complementary view. X-ray imager would specifically attempt to address the following
\begin{itemize}
    
\item Is the X-ray aurora driven by the
the same population of electrons
causing aurora in the VIS/UV?
\item What component of the X-ray
emission in the Polar Regions
comes from charge exchange
emission?
\item What is the correlation between
variability in the solar wind and the
spectral, morphological, and temporal
variations in X-ray aurora?
\item  Is there a diffuse X-ray emission from non-accelerated precipitating electrons in the absence of geomagnetic disturbances?
\end{itemize}

The instrument is an X-ray Pinhole camera which will
image the auroral arc and also measure the spectrum.
Auroral ovals are typically between 60$^{\circ}$ to 80$^{\circ}$ latitudes
and expand to lower latitudes during storms and substorms.
X-ray imager has a FOV of 130$^{\circ}$ to cover a 20$^{\circ}$latitude at any instant as the
satellite moves across the poles using two image plane detectors with individual pinholes and appropriate
mounting on the satellite. In addition to images, the X-ray camera will also provide the soft X-ray spectrum in the $\approx$ 0.3 to 3 keV range having a resolution of $\sim $ 100 eV at 1 keV. 
The imager will help map the auroral oval's boundaries and fine structures during storms/substorms. 
Two sets of identical imagers could be mounted on AuroraMag to capture the aurora. The X-ray imager will be developed as a heritage of the Auroral X-ray Imaging Spectrometer payload of the Disturbed and Quiet Time Ionosphere-Thermosphere System at High Altitudes (DISHA) mission.

\subsection{Electron and Proton particle sensor}

The high latitude of the Earth is bombarded by energetic particle precipitation.  To measure this particle detectors need to be deployed on the spacecraft platform. Recently, the Miniaturized Electron pRoton Telescope (MERiT) was flown on the 3U CubeSat, Compact Radiation belt Explorer, (CeREs)\citep{kanekal2019merit,kanekal2021dynamics}.  It is a low-mass, low-power, compact instrument. MERiT consists of front-end Avalanche Photodiodes (APDs) followed by a sequence of Solid-State Detectors (SSDs) enclosed within shielding made of Tungsten (W) and Aluminum (Al). The Avalanche Photodiodes measure low-energy electrons (5-200 keV), while Solid-State Detectors measure higher-energy electrons (1 MeV) with $30\%$ energy resolution. We proposed to utilize this detector to measure precipitating electrons and protons in the auroral region. 

The X-ray imaging will provide information on the energetic electrons that will produce X-rays. We need to depend on particle detectors (electron and proton) to get the complete details of precipitating electron energies. Therefore, by combining the in situ particle measurements with the X-ray measurements, one can also derive important information about the energy dependence of the generation and evolution of X-ray aurora and further identify the physical processes in underlying auroral features in greater detail.  The detectors will be mounted facing the zenith.

\subsection{Magnetometer}
\noindent
Magnetic field measurements with high accuracy are essential onboard AuroraMag to achieve identified science objectives. The in situ magnetic field measurements are crucial to study the magnetospheric and ionospheric current systems and deriving the pitch angle distribution of measured particles by MERit payload. Two identical fluxgate magnetometers (FGM) are proposed onboard each AurorMag spacecraft. They could be mounted on two opposite sides of the spacecraft body by using a deployable boom. The broad primary specifications of the FGM required for achieving the intended science objectives are listed in Table 3.

\begin{table}[h]
    \centering
    \begin{tabular}{c|c}\hline
      Parameter & Specs  \\\hline
      Range, Resolution, Accuracy & $\pm$ 60,000 nT, 1 nT, 2 nT \\
      (2 ranges)                  & $\pm$ 20,000 nT, 0.5 nT, 1 nT \\\hline
      Sampling rate & 8 vectors/s for FGM sensors, 16 bit/sample\\\hline
      Operating Temperature & Sensors and Electronics: - 20 deg C to + 55 deg C\\\hline
      Non-linearity & < 10 ppm \\\hline
      Total FGM Mass & 3 kg [Sensors: 0.5 kg; Electronics: 2.25 kg; Harness/Thermal: 0.25 kg]\\\hline
      Raw Power & 15.0 W \\\hline
      Size (Triaxial sensor) & 85 mm $\times$ 65 mm $\times$ 55 mm \\\hline
    \end{tabular}
    \caption{Specifications of the proposed FGM}
    \label{FGM-spec}
\end{table}

\noindent
The FGM sensors could be developed as a heritage of the magnetometer on ISRO's Adity-L1 spacecraft. Additional information on the magnetometer can be obtained by referring to \cite{yadav2018science}. 



\subsection{Electron Temperature  Analyser (ETA)} 
The Electron Temperature probe (ET probe) was initially designed by \cite{hirao1965rocket} which was later improved by \cite{hirao1970effect} in Japan. This ET probe was used in several rocket experiments in Japan ($\sim$50 flights) and also in India \citep{jain1981joule}. The probe was also used in HINOTORI \citep{oyama1991electron}, OHZORA \citep{oyama1987anisotropy}, and Exos-D \citep{abe1990measurements}. \\
This probe is one of the modifications of the resonance probe, in which the i-v characteristic curve is modified by superposing the RF voltage on the d-c biased probe. However, this instrument was not designed to have N$_e$ measurement capability. In this context,  \cite{oyama2015electron} proposed a modification to the basic ETA to enable the measurement of both N$_e$and N$_e$.  This instrument is compact, with low power consumption, and low data rate \cite{oyama2015electron}, and is suitable for smallsat missions.  For the AurorMag mission, we propose to have this Electron Temperature \& density Analyser (ETA), with the programmable capability to work in either N$_e$ and T$_e$ mode or only in Te$_e$. mode depending on the altitude. 

As mentioned earlier, the   T$_e$  is obtained by measuring the shift in the floating potential because of the applied RF signal, and N$_e$ measurement is done by considering the impedance effect on the probe, similar to the impedance probe principle, by detecting the upper hybrid resonance frequency. The details can be found in \cite{oyama2015electron}.
 
The proposed ETA sensor consists of two semicircular copper electrodes of  $\sim$150 mm in diameter and thickness of $\sim$2 mm with $\sim$10 mm gap between the two disks. In the T$_e$ mode, these two disks are electrically separated. When a sinusoidal voltage ($\sim$200 kHz) is added to the 
sweeping voltage of one disk, while the other is fed only with the DC sweeping voltage. By Subtracting these two signals using a differential amplifier, we can eliminate noise from the ambient plasma, including a variation of the floating potential of the satellite, and purely detect floating potential shift, which can be related to T$_e$. For the   N$_e$  measurements, we need to have the frequency sweep mode to obtain the Electron Density (N$_e$). Currently, the minimum frequency is 200 kHz, which means the lowest detection limit is $\sim$1000 electrons/cc. In an orbital altitude of $\sim$800-1000 km, we may expect the diurnal minimum of electron density could be as low as $\sim$200 electrons/cc. In these cases also, the T$_e$ mode will be able to provide temperature measurement. For example, the more recent Tatiana-2 mission was an experimental scientific education microsatellite in a circular near-polar, quasi-Sun-synchronous circular orbit at $\sim$830 km altitude which provided measurements of T$_e$ \citep{liu2015topside}.  Depending on the mission period and the selected altitude, we will decide whether ETA will have both N$_e$ and T$_e$ modes or only in Te$_e$ mode. The payload mass is expected to be $\sim$1-1.5 kg, and the power requirement will be within $\sim$5W.

\section{Spacecraft Bus requirements and challenges }

The payload specification, viewed from the perspective of science requirements, has been condensed into Table \ref{tab:ScReqMatrix}. Key payload requirements from the satellite bus are outlined in Table \ref{tab:SatSubSys}.

\begin{table}[h!]
    \centering
    \begin{tabular}{|p{3cm}|p{1.5cm}|p{1.5cm}|p{2cm}|p{2cm}|p{1cm}|p{2.5cm}|p{2cm}|}
         \hline
         \hline
         Payload & Power & Mass & Dimension &Pointing  & Data rate  & Duty Cycle & Total Data \\
         &in W & in kg & in mm & in deg & in Bytes/s & in seconds /orbit & in KB\\
         \hline
         Mag. fluxgate sensors & 15 & 0.5 & 85x65x55& NA&16&full orbit & 0.5 Gb\\
         Mag. Electronics & & 2.5 & 170x125x115 & &&& \\
         \hline
         ETA&5&1.5&15 cm diameter disc &Forward Hemisphere&TBD&<1000km in orbit& TBD\\
         \hline
         X-Ray Imager &20 &3 Kg& TBD &< 1&1.4e5&in orbit $\pm$60 degree and above &TBD\\
         \hline
         MERiT&0.5&1.2&40X40X15 &<1 &TBD&full orbit&TBD\\
         \hline
         \hline
    \end{tabular}
    \caption{Satellite Subsystems }
    \label{tab:SatSubSys}
\end{table}


Based on the table, it can be deduced that the overall payload mass is approximately 9 Kg. The spacecraft is anticipated to have dimensions of 50cm x 50cm x 50cm and a mass ranging between 30-35 KG. 

Achieving payload pointing requirements of $< 1$ degrees involves using actuators such as reaction wheels and magneto-torquers for momentum dumping and initial stumbling. Sensors utilized for pointing estimation include the Sun sensor, Earth Sensor, and star sensor. For control purposes, a MEMS-based IMU is also employed. The inclusion of a GNSS sensor provides location information in orbit for the magnetometer.


Assuming a total data budget requirement of approximately 2 GB per orbit for payloads and telemetry health parameters, with the proposed orbital inclination of $45^\circ$ and altitude of 400X10000 km, the orbital period is 3.44 hr. This leads to around 7  orbits per day, resulting in approximately 14 GB of data to be downloaded daily. Two ground stations, one at the Indian Institute of Space Science and Technology (IIST), Thiruvananthapuram, and another at ISRO Tracking and Ranging Centre (ISTRAC), are designated for telemetry/telecommand and data downloading requirements.  An elliptical orbit offers larger ground visibility at farther distances compared to a circular orbit. 




The Electrical Power System(EPS) in a small satellite comprises LiIon batteries and solar panels mounted on the PCBs. The total peak power requirement when all payloads are simultaneously active is around 50 W, but in practice, payloads are expected to be activated only when necessary.
The spacecraft bus is expected to consume about 30W of power, and the payloads would consume a peak power of 50W when all are active. However, after duty cycling the payloads, power consumption will be around 25-30W per orbit. For the 400 x 10000 km orbit, with a period of around 3.44 hours and an average eclipse period of barely 13 minutes, the spacecraft is expected to perform better in terms of power generation compared to a 1000 km circular orbit.

\section{Summary}
The solar wind interaction with the Earth's magnetosphere generates global response in currents, fields, and particles. The magnetosphere response is not always symmetric. There are various asymmetries in solar wind interaction with the Earth's magnetosphere like dawn-dusk, noon-midnight, hemispheric, etc. Understanding the asymmetries in solar wind-magnetosphere coupling is critical for space weather forecasting. The hemispherical asymmetric is one of the less-explored aspects of it. Investigating it using remote sensing and in situ, observations of both hemispheres simultaneously will provide invaluable data which was not available so far. The proposed mission concept aims to study this asymmetric response of the magnetosphere with twin smallsats observing southern and northern auroral ovals in x-rays along with in situ particle, magnetic field, and temperature measurements. The science objectives of the mission are well served by the optimized elliptical orbit with dimensions of 400 X 10,000 km.

Simultaneous data of imaging and in situ particle and magnetic field measurement will shed light on the degree of auroral and particle precipitation asymmetry and causes. Electron density and temperature measurements will probe the high-latitude heating of the ionosphere, cusp, FTEs, and its coupling to low-latitude.  Examining magnetospheric currents, such as field-aligned currents (FACs), and intensified ionospheric currents during space weather events is crucial for assessing the impact of space weather on the magnetosphere-ionosphere system. Additionally, conducting simultaneous observations in both hemispheres will aid in quantifying the asymmetry in solar energetic particle (SEP) propagation and understanding their connection to ground-level enhancements (GLEs). Further, collaborative observations involving operational space missions such as ERG, THEMIS, Cluster, and MMS, along with ground-based data from magnetometer and auroral imager networks, will contribute significantly to unraveling the complexities of wave-particle interaction and its role in aurora formation.

\section*{Acknowledgments}
The utilized data in this analysis is taken from the OMNI database. The data are publicly available at Coordinated Data Analysis Web (CDAWeb) https://cdaweb.gsfc.nasa.gov/
pub/data/wind/. We also
thank the Van Allen Probe mission team for making data available in the public domain.

\bibliographystyle{jasr-model5-names}
\biboptions{authoryear}
\bibliography{auroramag}

\end{document}